\documentclass[3p]{elsarticle}

\usepackage{amsmath,amsthm,verbatim,amssymb,amsfonts,graphicx,natbib}

\begin{document}

\begin{frontmatter}

\title{Numerical Simulation of the Hydrodynamical Combustion to Strange Quark Matter in the Trapped Neutrino Regime}
\author[cal]{Amir Ouyed \corref{cor2}}
\ead{ahouyedh@ucalgary.ca}

\author[cal]{Rachid Ouyed}
\author[usc]{Prashanth Jaikumar}

\address[cal]{Department of Physics and Astronomy, University of Calgary,
2500 University Drive NW, Calgary, Alberta, T2N 1N4, Canada}
\address[usc]{Department of Physics and Astronomy, California State University Long Beach,
1250 Bellflower Blvd., Long Beach, CA 90840 U.S.A}

\cortext[cor2]{Principal corresponding author}

\begin{abstract}
We simulate and study the microphysics of combustion (flame burning) of two flavored quark matter (u, d)  to three flavoured quark matter (u,d,s) in a trapped neutrino regime applicable to conditions prevailing in a hot proto-neutron star. The reaction-diffusion-advection equations for (u,d) to (u,d,s) combustion are coupled with neutrino transport,  which is modelled through a flux-limited diffusion scheme. The flame speed is proportional to initial lepton fraction because of the release of electron chemical potential as heat, and reaches a steady-state burning speed of (0.001-0.008)$c$. We find that the burning speed is ultimately driven by the neutrino pressure gradient, given that the pressure gradient induced by quarks is opposed by the pressure gradients induced by electrons. This suggests, somewhat counter-intuitively, that the pressure gradients that drive the interface are controlled primarily by leptonic weak decays rather than by the quark Equation of State (EOS). In other words, the effects of the leptonic weak interaction, including the corresponding weak decay rates and the EOS of electrons and neutrinos, are at least as important as the uncertainties related to the EOS of high density matter. We find that for baryon number densities $n_{\text{B}}\leq 0.35$ fm$^{-3}$, strong pressure gradients induced by leptonic weak decays drastically slow down the burning speed, which is thereafter controlled by the much slower burning process driven by backflowing downstream matter. We discuss the implications of our findings to proto-neutron stars. 
\end{abstract}


\end{frontmatter}

\section{Introduction}\label{introduction}

The Bodmer-Terazawa-Witten hypothesis (BTWH) \cite{bodmer1971collapsed,terazawa1979ins,witten1984cosmic} of absolutely stable quark matter made of up, down, and strange quarks ((u,d,s) matter) is of much astrophysical and fundamental interest.  One consequence of the BTWH is a large binding energy release associated with the conversion of a neutron star to a quark star. Since (u,d,s) matter has, by design of the BTWH, lower energy/baryon than hadronic matter, this conversion can release  up to $\approx 10^{53}$ ergs. The energetics and timescale of this conversion can be important for many high-energy astrophysical phenomena, e.g., Type II Supernovae, Gamma-Ray Bursts, and Superluminous Supernovae  \cite{ouyed2002quark,staff2008gamma,ouyed2013peculiar,shand2016quark,nakazato2013stellar,sagert2009signals,fischer2011core,anand1997burning,benvenuto1989evidence,benvenuto1989strange}. While energetically feasible and entropically favored, the dynamics of the conversion is quite complex and deserves further investigation.

But how does one describe the microscopic dynamics of the conversion of hadronic to quark matter? One way is by modeling the conversion as a combustion process with a well defined front, where the hadronic ``fuel'' is burnt into a (u,d,s) ``ash'' \cite{olinto1987conversion}; for a recent review of the literature about hadron-quark combustion, see Furusawa et al.\cite{furusawa2016hydrodynamical}.  The length of the reaction zone is determined by the leptonic and non-leptonic weak interaction and pressure gradients, and is of the order of centimeters \cite{niebergal2010numerical}. This length scale is six orders of magnitude smaller than the radius of a neutron star ($\sim 10$ km), which makes the problem computationally intensive. One is usually restricted to studying either the microphysics of the simulated ($\sim$ cm size) conversion zone \cite{olinto1987conversion,drago2007burning,niebergal2010numerical,furusawa2016hydrodynamical}, or reducing the combustion zone into a sharp discontinuity and propagating it as a turbulent deflagaration \cite{herzog2011three,manrique2015hydrodynamic}. In the latter scenario, typically, one assumes the combustion is a supersonic, shock-driven process (e.g. \cite{benvenuto1989strange}), or a subsonic, deflagration process (e.g. \cite{olinto1987conversion,furusawa2016hydrodynamical}), ignoring the fact that the burning speed must arise naturally from the nonlinear reactive-diffusive-advective processes that govern it. Niebergal et al. \cite{niebergal2010numerical}  (henceforth Paper I), modelled  the microphysics of the combustion process by numerically solving the reaction-diffusion-advection equations in the vicinity of the conversion front between two flavor (u,d) quark matter and  three flavor (u,d,s) matter. In this approach, nothing is assumed about the burning speed {\it a priori}, instead, the reaction-diffusion-advection equations are solved numerically including non-linear feedback effects between the weak decay rates, fluid transport, and the EOS. 
	
	An important result emerging from Paper I is that neutrino cooling (which was implemented in that work with an energy leakage scheme) generates thermal pressure gradients that slow down the burning speed. Furthermore, Paper I confirmed that for a given amount of neutrino cooling, the front could halt\footnote{We define the process of halting as in Paper I. Namely that pressure gradients induce a back-flow that advects the s-quarks away from  the interface and the downstream fuel. However, that does not mean that burning completely stops, given that the same back-flow causes downstream fuel to advect into the interface, converting it into deconfined (u,d,s) matter. In other words, halting means that the burning speed is nonzero, but order of magnitudes slower than if s-quarks were not advected backwards.}. Thus, for typical neutrino emission rates, the resulting cooling dominates the dynamics of the flame. Paper I clearly showed the significance of leptonic weak decays in the overall dynamics of the flame. Given that leptonic weak decays may be as important to the dynamics of the flame as the poorly constrained details of the EOS at high density, we present in this paper an improved study of the conversion dynamics that includes neutrino transport, going beyond a leakage prescription.
	
	 Specifically, we couple neutrinos to the hydrodynamics of the conversion, extending the microphysical approach of Paper I to the trapped lepton-rich regime. We implement neutrino transport through a flux-limited diffusion scheme and choose the physical conditions expected in proto-neutron stars as initial conditions, that is, high lepton fraction $Y_{\text{L}} \ge 0.2$ and $T=20$ MeV for initial tempcerature. Although some studies (\cite{vidana2005quark,lugones1998effect}) argued that neutrino trapping inhibited quark nucleation (and thus, subsequent combustion) in a proto-neutron star given that neutrinos raise the critical density required for deconfinement, there is a parameter window where quark nucleation is possible within the first seconds of a Supernova Type II explosion \cite{mintz2010thermal,sagert2009signals}. We use an updated version of the code originally used in Paper I called Burn-UD. Our results clearly show that leptonic weak decays are at least as important for the flame dynamics, if not more so, than the details of the high density EOS, because the pressure gradients generated by quarks are cancelled out by the pressure gradients generated by electrons, making the final burning speed a function of pressure gradients induced by neutrinos. Therefore, our finding suggests that studies of the combustion front require a treatment of leptonic weak decays, including neutrino transport. 
	  
	  	This paper is structured as follows: In Sec. \ref{formalism},  we list the system of equations for the combustion of (u,d) to (u,d,s). In Sec. \ref{microphysics},  we explain the physical effects that quarks, electrons, and neutrinos have on the hydrodynamics of the combustion flame. In Sec. \ref{results}, we present the numerical techniques that we used to solve the equations enlisted in Sec. \ref{formalism}; we also present the results of our simulations for the burning speed and discuss a semi-analytic model for the electron gradient induced instability. We discuss the astrophysical perspective of our results with concluding remarks in Sec. \ref{conclusion}.

\section{Combustion in lepton-rich matter}\label{formalism}

To ignite the combustion, we assume a  strange quark matter seed has already  been nucleated as an initial condition. As shown in Paper I, hadronic matter can begin to spontaneously combust into (u,d,s) matter provided a critical fraction of strange quarks diffuse across the interface (i.e, boundary of the seed) between hadronic and quark matter. Following a similar approximation done in Paper I, we  assume that hadrons have already dissolved into (u,d) matter and simulate the subsequent combustion of (u,d) to (u,d,s) matter. The  combustion process of (u,d) to (u,d,s) to quark matter in the lepton-rich trapped neutrino regime, where neutrino absorption is important, is driven by the following leptonic and non-leptonic weak interactions:

\begin{equation}\label{gamma1}
u + e^- \leftrightarrow s + \nu_{e}
\end{equation}
\begin{equation}\label{gamma2}
u + e^- \leftrightarrow d + \nu_{e}
\end{equation}
\begin{equation}\label{gamma3}
u + d \leftrightarrow u + s
\end{equation}

Reactions involving positrons and electron anti-neutrinos are not relevant to combustion in the lepton-rich regime in a proto-neutron star because they  are exponentially suppressed by the degenerate leptons. Although neutrinos of other flavors (e.g.$\mu, \tau$) arise through neutrino-pair bremsstrahlung,  their energy density is orders of magnitude smaller than electron neutrinos \cite{iwamoto1982neutrino}. Other weak interactions that are relevant are neutrino-quark scattering and  neutrino-electron scattering, which influence the transport of neutrinos across the interface.

The current version of Burn-UD solves reaction-diffusion-advection equations in 1D
\begin{equation}\label{oldEq} \partial U/\partial t=-\nabla F(U)+S(U)
\end{equation}
where U are the fluid variables as expressed in \cite{niebergal2010numerical}, with the addition of lepton number conservation. $F(U)$ are the advective-diffusive terms and $S(U)$ are the source (chemical reaction) terms ~\footnote{Paper I gives explicit forms for $F(U)$ and $S(U)$ in terms of relevant quark diffusion coefficients and weak decay rates.}. This work extends the system of equations 1-4 in Paper I by including the equations of neutrino transport:

\begin{equation}\label{eq5}
\dfrac{\partial n_{\nu_e}}{\partial t} + \nabla\cdot (\mathbf{v}n_{\nu_e}) +\nabla \cdot \left(\nabla D_{\nu_e} n_{\nu_e} \right) = \Gamma_1 + \Gamma_2
\end{equation}
\begin{equation}\label{eq6}
T\dfrac{\partial s}{\partial t} =\dfrac{d \epsilon_{\nu_e}}{d t}-\dfrac{d n_{\nu_e}}{d t}\mu_{\nu_e},
\end{equation} 

where $n_{\nu_e}$ is the number density of electron neutrinos,  $\epsilon_{\nu_e}$ is their internal energy density given as a function of $n_{\nu_e}$, $\partial s$/$\partial t$ is the entropy density change due to  electron neutrinos, $\mu_{\nu_e}$ is neutrino chemical potential, $T$ is temperature, and $\Gamma_1,\Gamma_2$ are reaction rates for \eqref{gamma1} and \eqref{gamma2} which are taken from \cite{anand1997burning}. $D_{\nu_e}$  is flux limited in such a way that neutrinos do not travel faster than the speed of light. Effectively,  $D_{\nu_e}= c \Lambda/\lambda_{\nu_e}$,where $\Lambda$ is a function given by  equation (28) in Levermore et al. \cite{levermore1981flux}, and $\lambda_{\nu_e}$ is the mean free path for neutrinos. We use analytic expressions for  $\lambda_{\nu_e}$ originally derived by \cite{iwamoto1982neutrino}. Since  neutrinos are degenerate, we may reasonably assume that only neutrinos with energy $E_{\nu_e}=\mu_{\nu_e}$,  are scattered or absorbed at the interface. We close the above equations with the thermodynamic Bag Model EOS for  quark matter as elaborated in Paper I, where hadrons were treated as dissolved two-flavor quark matter.

\subsection{Hydrodynamical Effects}\label{microphysics}

\begin{figure}[t!]
\centering 
\includegraphics[width= 0.85\textwidth]{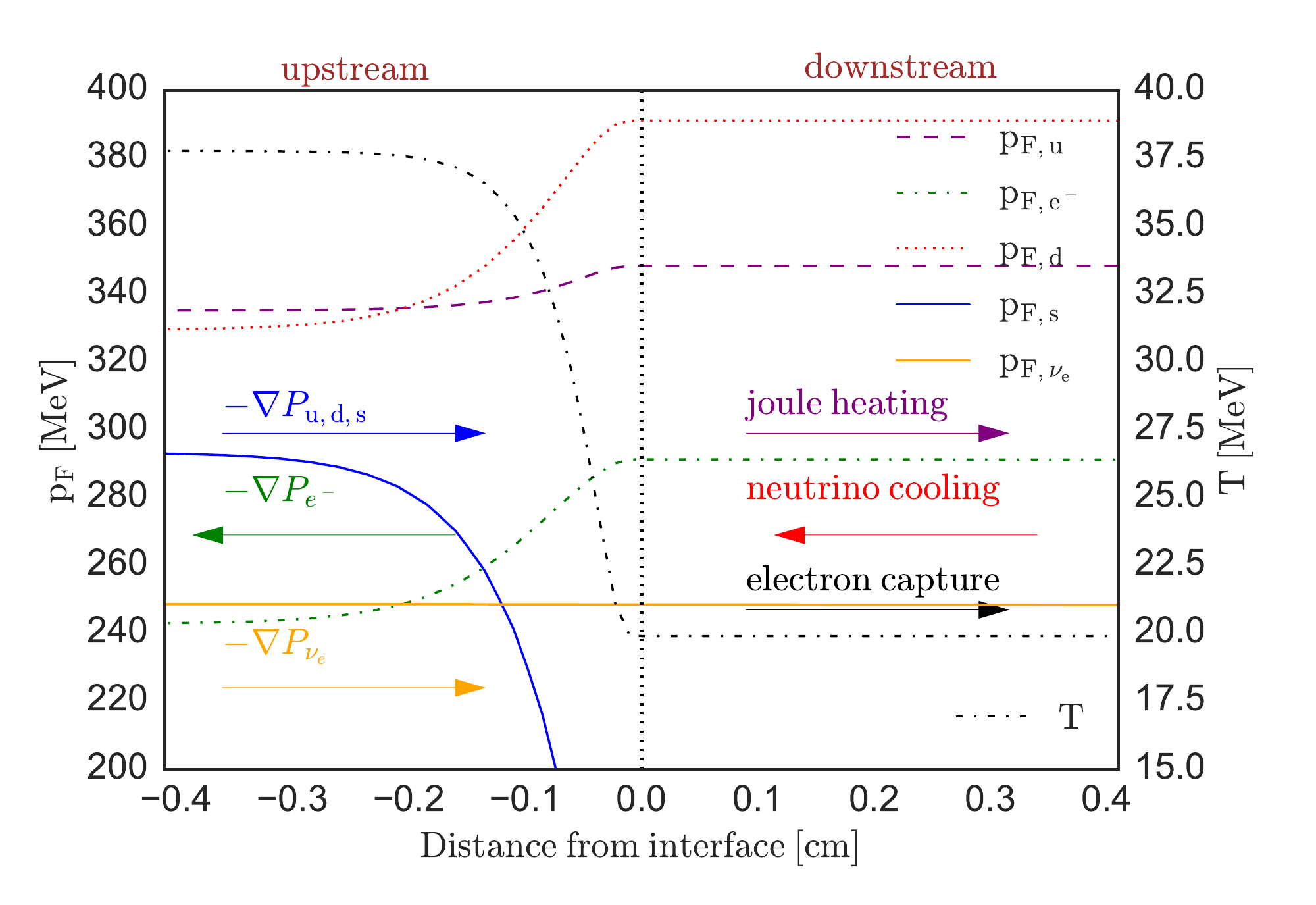}
\includegraphics[width= 0.85\textwidth]{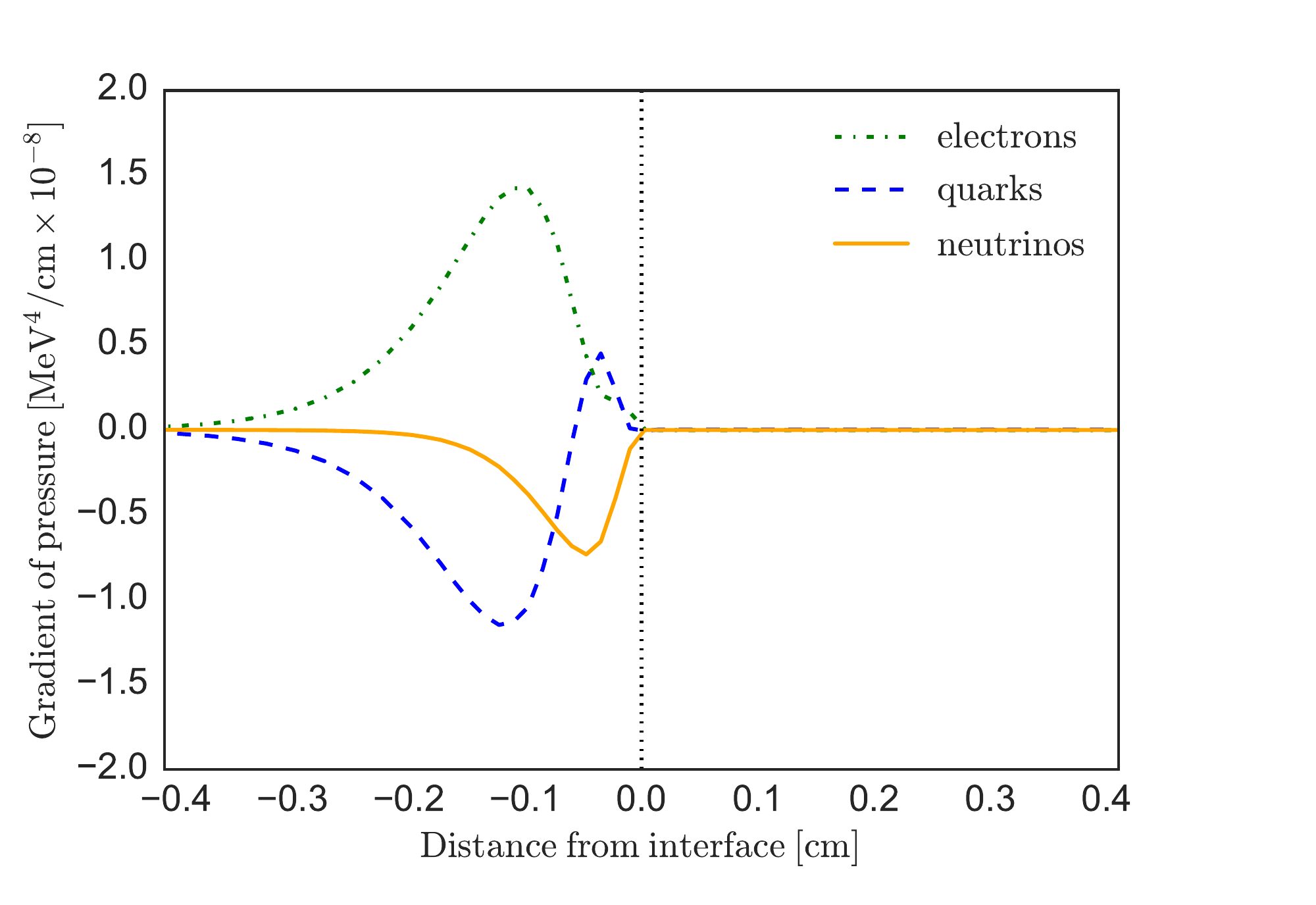}

  \caption{\textbf{Upper Panel}: Simulation snapshot of the burning interface. $p_{F,i}$ are the Fermi momenta for particles i, and $T$ is the temperature. In both panels, the interface lies at position zero depicted by the vertical line.  The arrows represent the directions of the force vectors and their labels depict the processes that caused them. Upstream is the side behind (left side of the vertical line) the interface, and downstream is the side in front of it (right side of the vertical line). 
   \textbf{Lower Panel}: The pressure gradients for the leptons and quarks shown in the upper panel.}
    
  \label{zoner}  \label{pgradient}
\end{figure}

In the BTWH, combustion of hadrons implies transport of strange quarks into hadronic matter, propagating the conversion. As noted in Paper I, the transport of strange quarks will be affected by pressure gradients that determine flow velocities. In the hydrodynamic approach, there are many forces acting upon the combustion front in contrast to the simplified approach where diffusion caused by the strange quark's density gradient is the sole mechanism of transport \cite{olinto1987conversion}. 
The hydrodynamic forces are related to pressure gradients and they in turn determine the speed of the burning interface (lower panel in Fig. \ref{zoner}). The potential sources of pressure gradients can be roughly divided into
those coming from concentration gradients of quarks and leptons (neutrinos and electrons), and those coming from temperature gradients induced by electron capture, neutrino diffusion and neutrino emission.

One can visualize the interaction between all these processes at the interface as a ``tug-of-war" between force vectors that point in the upstream or downstream direction (upper panel in Fig. \ref{zoner}).  We term as ``enhancing" the factors that cause a force in the downstream direction, accelerating the s-quarks (and hence combustion) through the interface. We term as ``quenching" the factors that create a force vector in the upstream direction, slowing down the transport of s-quarks into the interface. The relation of the microphysics to the dynamics of the interface determines the enhancing and quenching factors. For example, the factors that enhance combustion are: 

\begin{figure}[t!]
\centering 

\includegraphics[width= 0.55\textwidth]{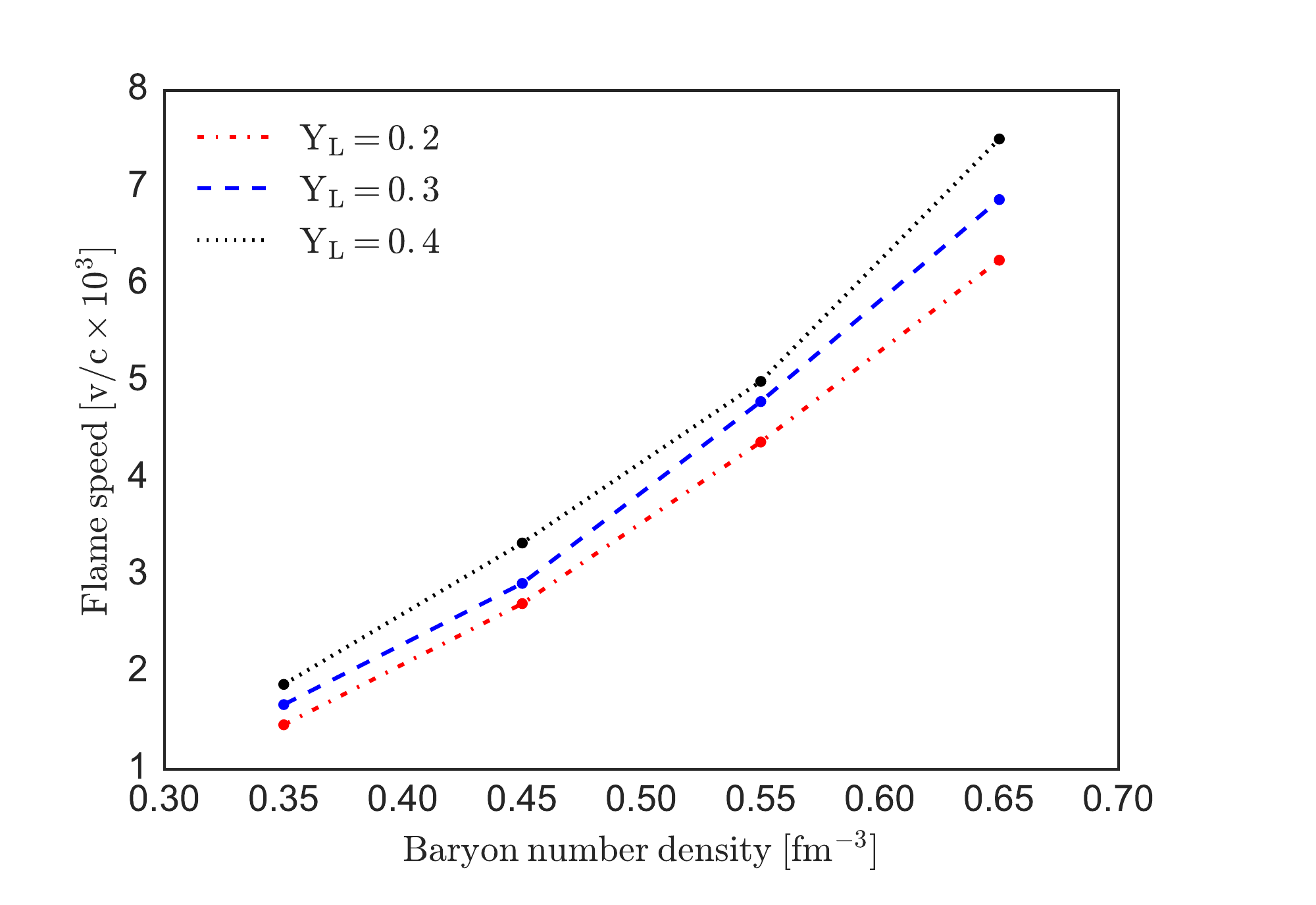}
\includegraphics[width=  0.55\textwidth]{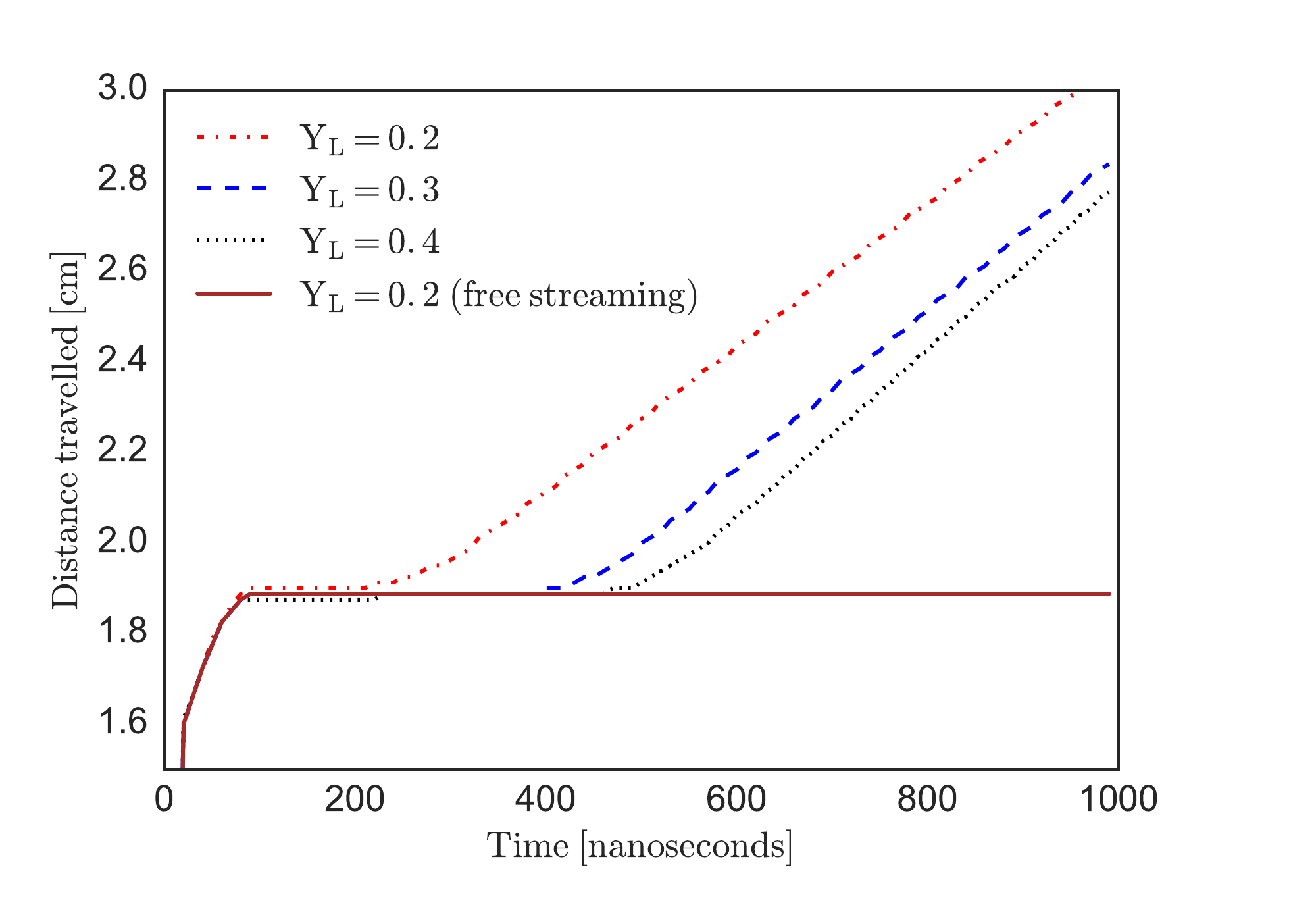}
\includegraphics[width=0.55\textwidth]{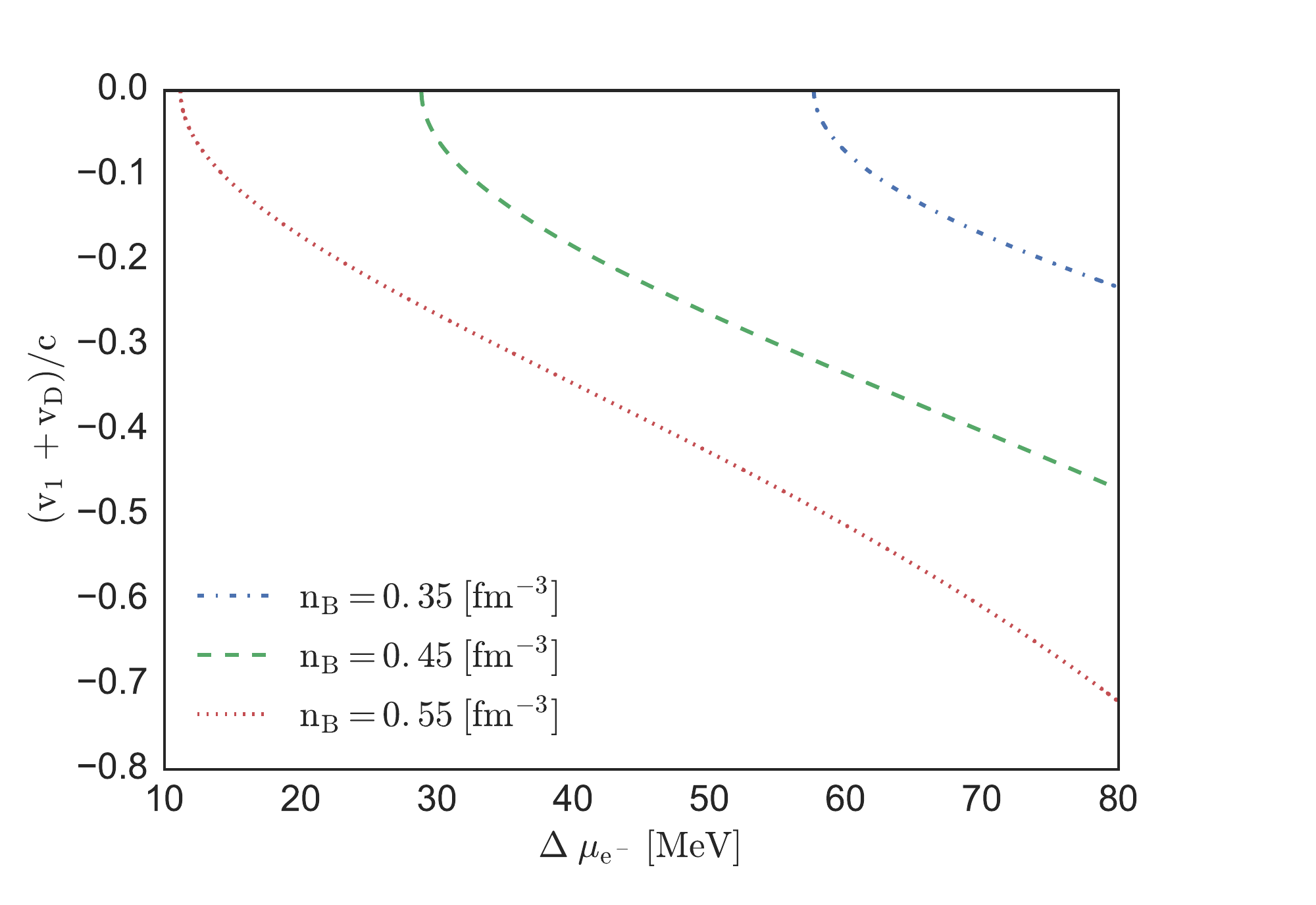}

\caption{\textbf{Upper Panel}: Burning speed as a function of baryon number density (n$_{B}$) and  initial lepton  fraction ($Y_L$) for the trapped neutrino regime. 
\textbf{Middle Panel}: Interface position as a function of time (in nanoseconds; in our simulations, a nanosecond corresponds to $\sim 10^4$ timesteps) when n$_{B}=0.35$ fm$^{-3}$ for the trapped and the free streaming regime ($Y_L=0.2$; solid line). Notice how the free streaming regime leads to zero velocities for the extent of the simulations.
\textbf{Lower Panel}: Semi-analytic solutions for quenching due to the deleptonization instability for different  baryon number densities $n_B$.  A negative value  of  $v_1+v_D$ implies the interface has halted, and instead combustion happens at a much slower rate, because burning is only driven by the back-flow of fuel.}\label{corepos}\label{vel2}\label{toy}
\end{figure}

\begin{figure}[t!]
\centering

\includegraphics[width= 0.55\textwidth]{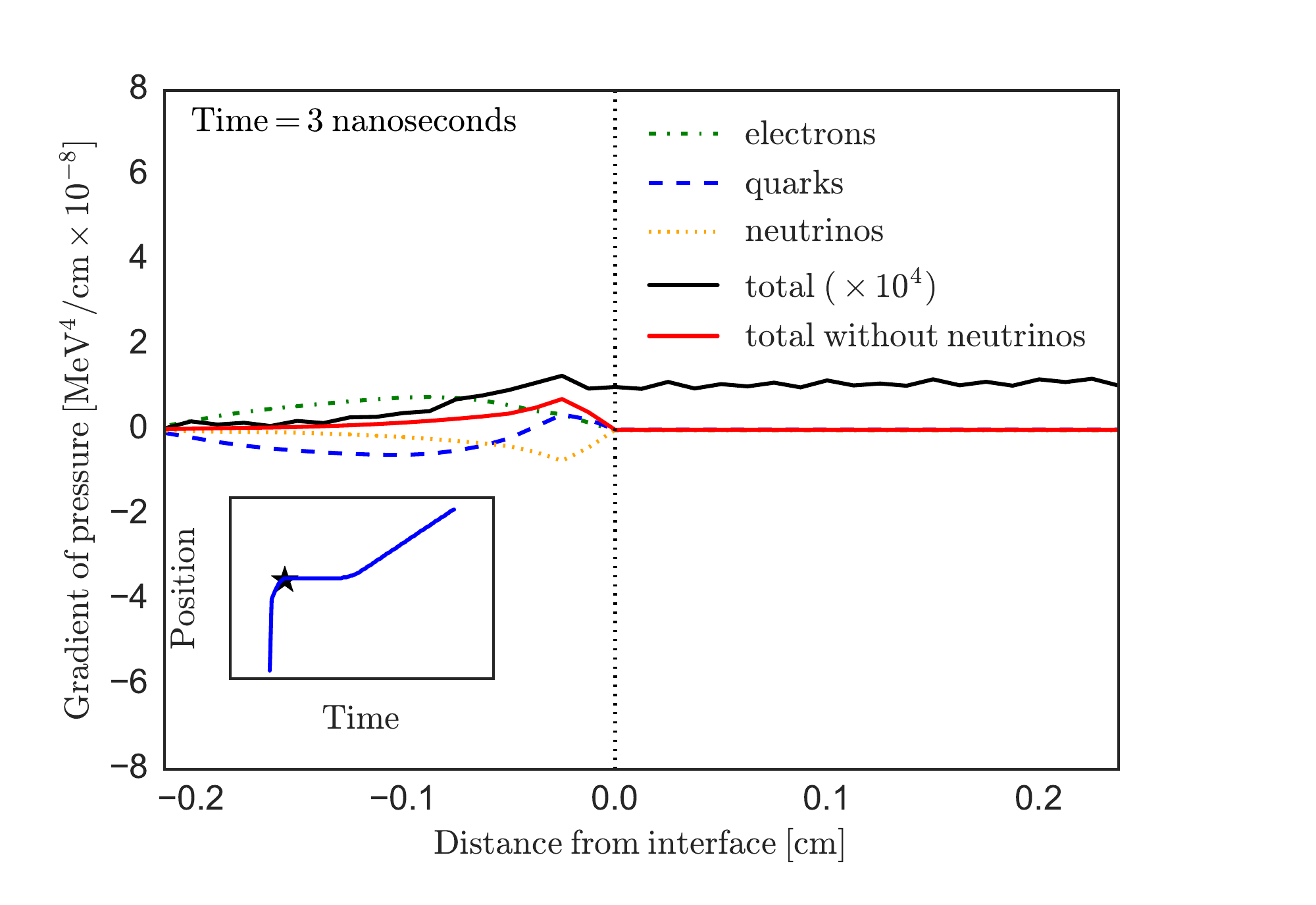}
\includegraphics[width=0.55\textwidth]{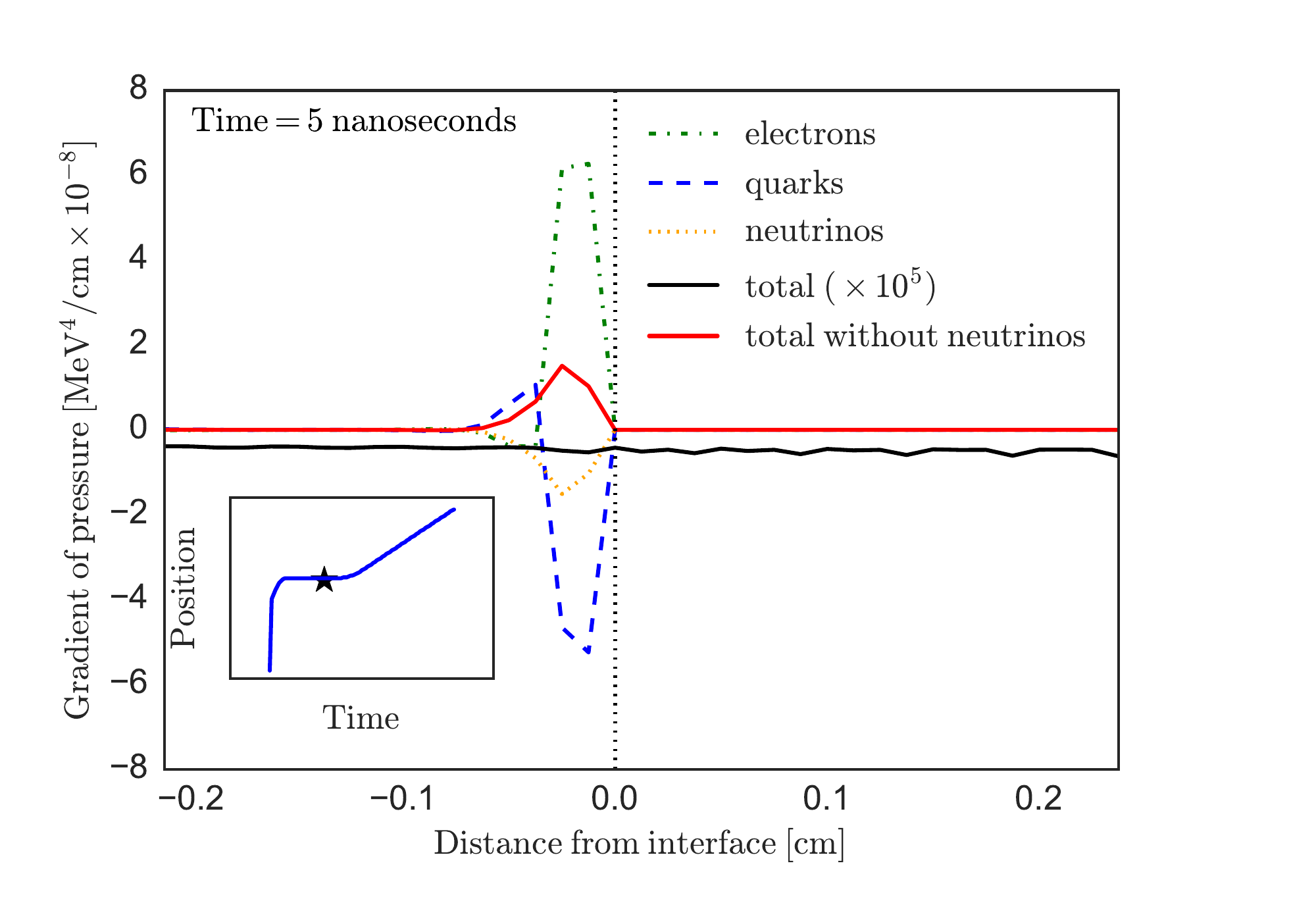}
\includegraphics[width=0.55\textwidth]{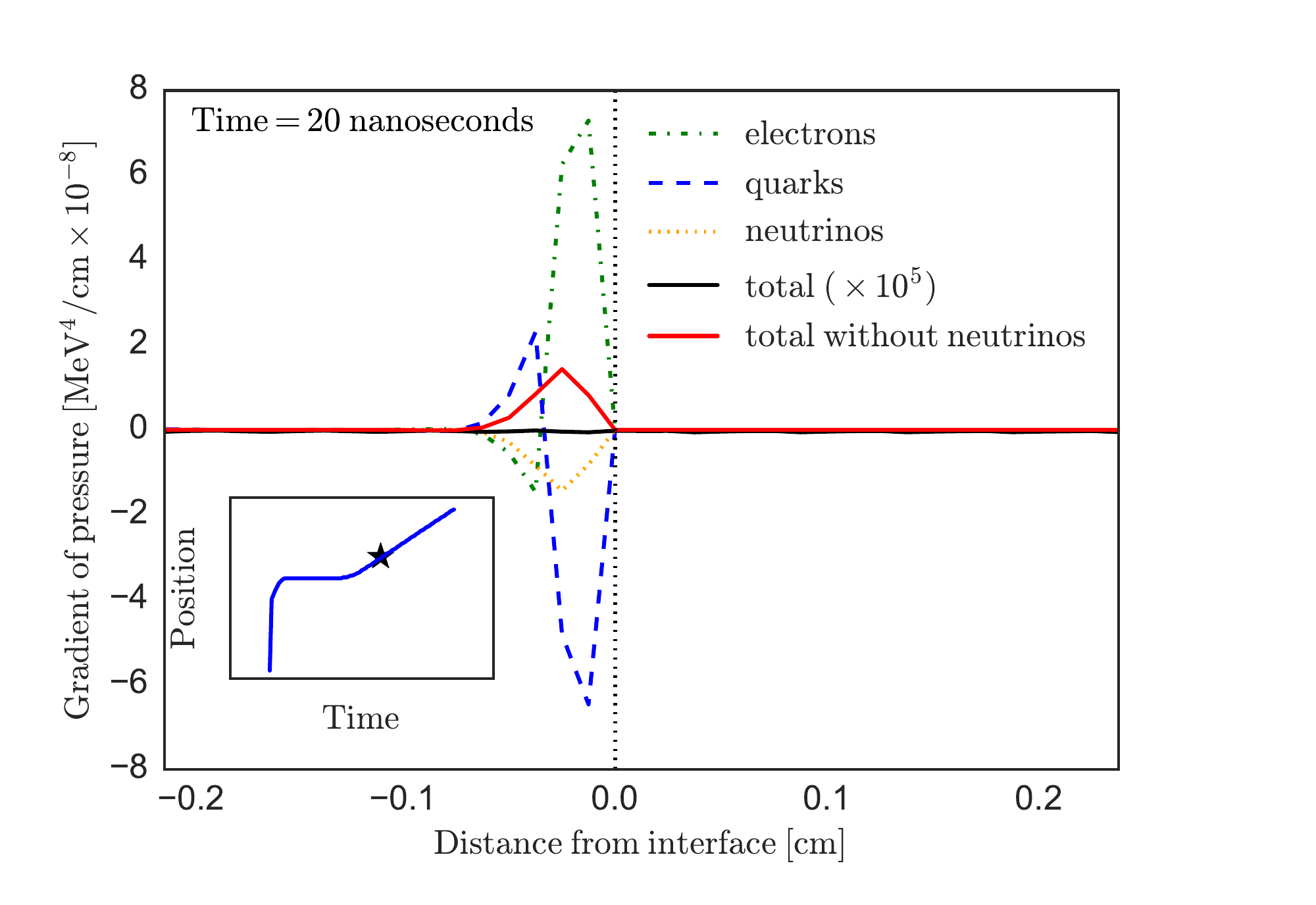}
\caption{Pressure gradients versus interface position at different times, for the case of $n_{\rm{B}}=0.35$ fm$^{-3}$ and $Y_L=0.3$.  The total pressure gradient induced by both leptons and quarks is labeled as ``total". The sum of the pressure gradients of quarks and electrons only is labeled as ``total without neutrinos". The line for ``total", not to scale, is multiplied by the factor shown in parenthesis, given that the sum of all pressure gradients tends to cancel out. We plot as an inset position versus time of the interface, and use a black star to signal the position and time for the snapshot.} \label{sequence}
\end{figure}

\begin{enumerate}

	\item Flavor equilibration: The population of the strange quark energy levels decreases the energy/baryon, with some of that energy deposited as heat, raising the temperature of the quark matter behind the interface. The (u,d,s) EOS has a $T^4$ (see equation (5) in Paper I) dependence on temperature, therefore the temperature gradient will lead to a negative pressure gradient ($dP_{\rm quark}/dx < 0$, where $P$ is pressure and $x$ is position) enhancing the transport of strange quarks across the interface. Although the quarks have overall a  negative pressure gradient, sometimes there are positive pressure gradient wiggles, such as the one in the lower panel of Fig. \ref{zoner} appearing just behind the interface. This has to do with nonlinearities in the quark EOS, which has the form of $P=a \sum_{i=u,d,s}\mu_i^4+ b\sum_{i=u,d,s}\mu_i^2 T^2 + c T^4$.  The first component softens as quarks flavor-equilibrate, while the second and third components stiffen due to the temperature increase triggered by equilibration.  Usually the gradients of the second and third component dominate, hence the overall negative pressure gradient, but sometimes the gradient of the first component causes positive pressure gradients. We note the induced wiggles are higher order terms in an otherwise overall negative pressure gradient.
	
	\item Electron capture: This increases the temperature gradient behind the interface by raising the entropy density  $\Delta s=-\dfrac{\partial n_e}{\partial t}\mu_e-\dfrac{\partial n_{\nu_e}}{\partial t}\mu_{\nu_e}$ (see equation (6) in Paper I), where  $n_e$ is the electron number density, and $\mu_e$ is electron chemical potential.  Although for every electron capture, a neutrino is produced, $\mu_e$ is generally greater than $\mu_{\nu_e}$, so there is  net heating involved. This enhances the pressure in the upstream region.
		
	\item Neutrino pressure: In the trapped neutrino regime (hereafter trapped regime), this enhances burning by exerting a negative pressure gradient. This comes from the negative density gradient of neutrinos across the interface as a result of reactions \eqref{gamma1} and  \eqref{gamma2} creating neutrinos behind the interface.
	
	\item Loss of lepton number: a lepton-rich fluid parcel behind the interface lowers its lepton content over time due to neutrino diffusion, leading to Joule Heating \cite{burrows1986birth} by increasing the entropy $\biggl(\dfrac{\partial Y_L}{\partial t}\mu_{\nu_e}>0\biggr)$. Lepton loss is related to point 2, in the sense that it is electron capture that transforms electrons to more mobile neutrinos that eventually leave the interface. However, while electron capture tends to raise the temperature behind the interface, lepton loss becomes relevant only  near the interface, where the neutrino gradient is large and therefore  $\dfrac{\partial Y_L}{\partial t} \propto D_{\nu_e} \dfrac{\partial^2 Y_L}{\partial x^2}$, where $x$ is position,  becomes large.

	\end{enumerate}
	
\noindent The quenching factors are: 
	\begin{enumerate}
	
	\item Electron pressure:  There is a positive electron gradient across the interface due to Eqs.  \eqref{gamma1} and  \eqref{gamma2}.  I.e. electron capture depletes electrons behind the interface,  forming  a large dip in the electron number density behind the interface (see upper panel Fig.  \ref{zoner}). As the electron EOS is proportional to $n_e^{4/3}$, this leads to a positive pressure gradient that induces a force vector pointing in the upstream direction. 
	
	\item  Neutrino cooling: As neutrinos escape from the burning front, they drain energy from the fuel, reducing the temperature. As the front propagates and emits neutrinos, it will leave behind a positive temperature gradient which leads to a negative force vector pointing in the upstream direction, slowing down the burning. The effect on (u,d) to (u,d,s) combustion was the focus of Paper I. 

    \end{enumerate}

Another important effect of the hydrodynamics on the combustion zone is that the hydrodynamics affect the width of the reaction zone.  A typical formula used to estimate the width of the reaction zone is $l \sim \sqrt{D \tau}$, where  $\tau \sim 10^{-8}$s \cite{furusawa2016hydrodynamical} is the timescale of the nonleptonic process  (Eq. \eqref{gamma3}) \cite{alford2015phase} and $D$ is the quark diffusion coefficient. However this formula assumes that the only important process is diffusion, and therefore disregards hydrodynamics. Using that formula will give a small width of  $l \sim \sqrt{D \tau} = 10^{-5}$cm - $10^{-4}$ cm which is various orders of magnitude lower than the numerically calculated length $\sim 1$ cm, because advection tends to widen the flame width given that the fluid velocities transport the u and d quarks through a centimeter for $\sim 10^{-8}$s before they decay into s quarks.

\section{Numerical Simulations and Results}\label{results}

The variables in the equations of hydrodynamical combustion (Eq. \eqref{oldEq}) are solved for numerically using an upwinded, third order spatial discretization for the advection terms, while the diffusion terms are discretized with second order central scheme then these equations are integrated temporally with a fourth order, Runge Kutta algorithm. Since neutrino transport happens on a much faster timescale than the hydrodynamics, we solve the hydrodynamics explicitly (Eq. \eqref{oldEq}) while coupling the  neutrino transport implicitly (Eq. \eqref{eq5}). We use  first order operator splitting to couple the explicitly evaluated reaction-diffusion-advection equations (equations 1-4 in Paper I) with the implicitly solved neutrino transport equations (Eq. \eqref{eq5}-\eqref{eq6}). Eq. \eqref{eq5} is spatially discretized with second order central scheme, and  is implicitly time  integrated with a first order, backward Euler scheme. Eq. \eqref{eq6} is computed from  the $n_{\nu_e}$ value evaluated from Eq. \eqref{eq5}. 
The large pressure gradients due to micro-physical effects appear in a very narrow zone, so large resolutions are required to resolve the effects. For the combustion front module,  we used 48000 zones for a 600 cm grid, and a timestep $dt$ that satisfies $c dt/dx$=0.4, where $c$ is the speed of light and  $dx=0.0125$ cm is the length of the zone. To solve the linear system of the neutrino transport equations generated by the Backward Euler scheme, we used a java implementation of LAPACK \cite{MTJ}.

\subsection{Steady state burning speed}

We run Burn-UD for an initial temperature $T=20$ MeV and the densities and lepton fractions depicted in the upper panel of  Fig. \ref{vel2}. Once the initial lepton fraction $Y_{L}$ is fixed, beta-equilibrium for two-flavor quark-matter determines the initial electron and neutrino fractions. The upper panel of Fig. \ref{vel2} shows that increasing $Y_{L}$ or increasing baryon number density n$_\text{B}$ both have the similar effect of increasing the burning speed for the trapped neutrino  case. This dependence of the burning speed on $Y_{L}$  is expected because of the enhancing effect of electron captures, as explained in Sec. \ref{microphysics}. Higher densities lead to higher burning speeds simply because there is more  (u,d)``fuel'' to burn, which leads to higher temperatures behind the front.

\subsection{Deleptonization instability}

Fig. \ref{sequence} shows a sequence of time snapshots of the pressure gradients across the interface for $n_B=0.35$ fm$^{-3}$ and $Y_L=0.3$. From Fig. \ref{sequence}, we can derive the following observations about the pressure gradients: 
\begin{enumerate}

\item{The pressure gradients of leptons and quarks tend to cancel each other out, as the total pressure gradient is about four orders of magnitude smaller than the pressure gradient of quarks or leptons.}
\item{Although the final, steady burning speeds are nonzero (see middle panel of Fig. \ref{corepos}) the burning front halts temporally. The insets in Fig. \ref{sequence} show a zero velocity plateau in position versus time. The halting corresponds to a positive total pressure gradient that induces negative fluid velocities and opposes the movement of the interface, as shown in the snapshot for time = 2.7 nanoseconds. However later on, the total pressure gradient becomes negative, reviving the burning front.   }
\item{The neutrino pressure gradient is the component that gives the final ``push" to the interface, and therefore determines whether the burning front halts or not. Without the pressure gradient induced by neutrinos, the total pressure gradient near the interface would remain positive (as shown by the red, solid lines in Fig. \ref{sequence}), and therefore the burning front would remain halted.} 
 
\end{enumerate}

The behavior of the burning front can be summarized as follows: reactions \eqref{gamma1} and \eqref{gamma2} deplete the upstream region of electrons, causing a large electron gradient that induces a force vector in the upstream direction (Fig. \ref{zoner} and Fig. \ref{sequence}). At some point, this force vector is large enough to induce fluid velocities in the upstream direction that oppose the diffusion of s-quarks into the downstream direction, halting the  burning. However,  the enhancing processes revive the burning front (see Sec. \ref{microphysics} for a discussion about enhancing processes). The burning front is revived because neutrinos exert a pressure gradient that opposes the quenching effects of the electron pressure gradient  (Fig. \ref{pgradient}) and they also deposit thermal energy through Joule heating (point 4 of the enhancing processes in Sec. \ref{microphysics}), increasing the pressure in the upstream region.  Eventually, the enhancing processes overwhelm the electron pressure quenching, reviving the burning. 

In order to prove that the neutrinos are ultimately decisive in reviving the burning front, we ran simulations where we turned off neutrino effects, such as neutrino pressure, neutrino energy transport, and Joule heating. This emulates roughly the free-streaming case scenario, in the sense that neutrinos streaming out of the interface would not deposit energy or momentum in the reaction-zone. Free streaming is also a plausible approximation for $n_\text{B}\leq 0.35$ fm$^{-3}$, given that neutrino mean free path is roughly of the order of one centimeter, which is the width of the reaction zone. When we turn off neutrino effects, the burning front, once halted, remains so for the length of the simulation (middle panel in Fig. \ref{corepos}).  This proves that neutrinos, and thus, leptonic weak-decay physics, are ultimately decisive in the dynamics of the flame.

What would happen in a multidimensional scenario? Anisotropies along the interface, such as inhomogeneities in electron number density, would lead to uneven halting among different parts of the interface. In other words, some parts of the interface will keep propagating while others will stop. Furthermore, if a part of the interface halts, it will recede inwards, in the upstream direction, given that halting is caused by electron pressure gradients inducing negative, inward velocities. This will result in parts of the interface moving in opposite directions, in other words, in shearing. A sheared interface will have a larger surface area which leads to a faster burning rate, as the interface can absorb more fuel per unit of  time. However, there will be also a stabilization effect, given that the s-quark diffusion tends to wash out anisotropies in the interface. We argue that the interplay between stabilization and shearing could lead to a hydrodynamic instability. As in Paper I, we refer to this effect as a ``deleptonization instability", given that the halting phenomenon is due to the electron dip caused by leptons diffusing out of the upstream region  (see Sec. \ref{microphysics}). Although the leptons remain trapped in the proto-neutron-star/quark-star, there is a local deleptonization experienced by the upstream side of the interface.

	\subsection{Semi-analytic model of deleptonization instability }\label{toySection}
	
	The ``deleptonization instability"  can be captured with a semi-analytic, simplified model based on the hydrodynamic jump conditions. Although the pressure gradients that induce this instability appear in a very narrow zone (see Fig. \ref{pgradient}) this treatment provides a simple, alternate way of exploring the main physics behind the quenching due to an electron pressure gradient. 
	
	Let us assume both the upstream and downstream regions are in beta-equilibrium, given that only in a narrow $\sim$ cm zone are quark abundances changing (see upper panel in Fig. \ref{zoner}). Furthermore we assume all quarks are mass-less and ensure charge neutrality. One can deduce the quark abundances given an electron number density $n_e$ and a baryon number density ${n_B}_i$ where $i$=1 is downstream (i.e. two-flavor bag model) and $i$=2 is upstream  (i.e. three-flavor bag model). The corresponding expressions are:
	
	\begin{equation}
	{n_s}_2=\frac{{n_B}_2-{n_u}_2}{2}
	\end{equation}
\begin{equation}
	{n_u}_2={n_B}_2+{n_e}_2
\end{equation}	
	\begin{equation}
	{n_d}_2=3{n_B}_2-{n_s}_2-{n_u}_2
\end{equation}

where the subscript 2 labels the upstream variables. In the downstream case, there are no s-quarks, so:
\begin{equation}\label{2}
	{n_u}_1={n_B}_1+{n_e}_1
\end{equation}	
	\begin{equation}
	{n_d}_1=3{n_B}_1-{n_u}_1
\end{equation}
 
We equate ${n_B}_1=n_B$ and ${n_B}_2=0.95 n_B$, where $n_B$ is a parameter. The jump conditions then become (see Paper I):

\begin{equation}
({\mu_s}_2^3+{\mu_d}_2^3+{\mu_u}_2^3)v_2=({\mu_d}_1^3+{\mu_u}_1^3)v_1
\end{equation}
\begin{equation}
h_2 v_2^2+P_2=h_1 v_1^2+P_1
\end{equation}

We assume that neutrinos decouple from matter near the interface, as explained at the end of Sec. \ref{microphysics}. Thus, we do not include enthalpy or pressure terms for neutrinos in the jump conditions.

Once the front is halted, the transport of s-quarks is only due to diffusion and advection, and a solution for the halting due to the electron-gradient exists when $v_1+v_D\le 0$, where $v_D\sim D/l$, where $D$ is the diffusion coefficient for quarks which is always around $D \sim 0.1$ cm$^2$/s. Here, $l$ is the width of the interface which is of the order of a centimeter. This condition is equivalent to stating that negative fluid velocities must prevent s-quarks from diffusing into the downstream direction to quench the burning. 

We fix ${n_e}_1=0. 4 n_B$, $T_1=18$ MeV, $T_2=38$ MeV and we vary the parameters  $n_B$ and  ${n_e}_2$ to calculate the quantity $v_1+v_D$ (lower panel in Fig. \ref{toy}). The variation of the parameter  ${n_e}_2$ represents neutrino transport, as electrons in the upstream region are depleted as time progresses due to electrons ``converting" into neutrinos and diffusing out of the interface. We plot  $v_1+v_D$  as a function of  $\Delta \mu_e={\mu_e}_1-{\mu_e}_2$ (lower panel in Fig. \ref{toy}). We find that for  $\Delta\mu_e \ge 10$ MeV, there are quenching solutions. This range is expected in the simulations (see upper panel in Fig. \ref{zoner}). A  negative $v_1+v_D$ leads to the same effect that is depicted in the middle panel of Fig. \ref{corepos}, where downstream back-flows essentially halt the interface. 

The quenching of the burning front by back-flows is superficially similar to the dramatic slow down of the burning-speed triggered by the so called Coll's condition in other papers (e.g. \cite{drago2015combustion}, \cite{herzog2011three}). Coll's condition states that hydrodynamic combustion will halt when the energy density of fuel is less than the energy density of ash for a given pressure and volume. Drago et al. 
\cite{drago2015combustion} argued that once  Coll's condition is violated at a specific critical density, hydrodynamic combustion shuts off in the outer layers of the compact star, which renders burning as purely diffusive.  However, in our case, fuel back-flows prevent s-quarks from diffusing downstream, which slows down the burning considerably. Both approaches lead to dramatic slow down of burning speed, but the physical causes are different and therefore the timescales for combustion of the whole compact star will differ between the two approaches. In our case, Coll's condition does not apply for the reasons stated in \cite{furusawa2016hydrodynamical}, namely that combustion is not a process in mechanical equilibrium, where the ash and the fuel have the same pressures and volumes, and there is no energy barrier quenching the burning, given that enough s-quarks are available.

\section{Conclusions}\label{conclusion}

			We solved the reaction-diffusion-advection equations for the combustion of (u,d) to (u,d,s) matter in the trapped  regime for conditions applicable to proto-neutron stars, treating neutrino transport in a flux-limited approach. Our main finding is that the combustion dynamics, including the burning speed, are dominated by the leptonic weak interaction effects rather than the details of the  quark EOS. This is because the electron pressure gradient cancels out the pressure gradient generated by quarks, making the pressure gradient generated by neutrinos the final decisive component for the burning speed. Since the burning speed is extremely sensitive to leptonic weak decay micro-physics, any study of the hydrodynamics of the combustion processn that does not incorporate the microphysics of lepton decay will be inadequate. Furthermore, the leptonic weak decay reaction rates and the EOS of neutrinos and electrons, in addition to the evolution of entropy, are nonlinearly coupled. This creates feedback effects that can either slow down the burning front or accelerate it. However, we find that burning speeds are bounded by the lepton fraction to less than $\sim$ 0.01c (see upper panel in Fig. \ref{corepos}). Electron pressure gradients can wrinkle the interface, which could enhance the burning speed, possibly leading to detonation. Paper I already predicted ``deleptonization instability" in the cold neutron star, and in this work we comfirm a similar phenomenon in the hot, proto-neutron star case. 
			
			A deficiency in our study is the lack of a hadronic EOS, because our work assumes hadrons have dissolved already into (u,d) matter.   If  a realistic hadronic EOS is considered then its effects on the flame dynamics depend on whether the phase transition is discontinous or mixed and smooth. If the hadron-quark transition appears discontinously, it could trigger a shock given a sudden change in pressure, which could lead to detonation. However if the phase-transition is smooth, such as in the case where it appears through a mixed phase, it would likely lead to similar results to this paper, given that the pressure gradient induced by the change of hadronic to quark EOS would not be drastic enough to affect the dynamics of the combustion front. Whether the transition is discontinous or smooth, depends on the surface tension of quark matter, where a high surface tension would hamper a mixed phase. Finally, our study assumes that (u,d,s) matter is stable at zero pressure, yet there are quark EOSs where stable quark matter only appears at high pressures; these EOSs lead to the creation of a hybrid star. It would be interesting to explore alternate quark EOSs in the context of combustion, that include more realistic physics, such as chiral symmetry breaking.

			Although, we focused on the microphysics of the flame, which has a length-scale of  a centimeter, we can effectively explore and model much larger scales relevant to the proto-neutron star, by varying the baryon number density, $n_\text{B}$. The finding that for $n_\text{B}<0.35$ fm$^{-3}$ the flame is halted, suggests a contained quark core of a few kilometers in the proto-neutron star case.
		Nevertheless, the core size can grow, even if the front is halted, at  a slow rate defined by the slower, negative upstream velocities (i.e. back flow) which advect fuel into the quark core, where they would transform to (u,d,s) matter by the BTWH. Given that our work suggests that the effects of electron pressure and neutrino emission can reduce the burning speed by orders of magnitude, the calculation of the conversion timescale of the whole compact star requires the inclusion of lepton microphysics and neutrino transport. However, an estimate of the conversion timescale can only be quantified with a global, hydrodynamic simulation of the whole compact star, which is beyond the scope of this paper.

\section*{Acknowledgements}
 R.O. is funded by the Natural Sciences and Engineering Research Council of Canada. P.J. is supported by the U.S. National Science Foundation under Grant No. PHY 1608959. We thank N. Koning, Z. Shand, and L. Welbanks  for discussions.

\section*{\refname}
\bibliographystyle{elsarticle-num}
\bibliography{Draft2Bib}

\end{document}